# A note and a correction on measuring cognitive distance in multiple dimensions

Ronald Rousseau [a,b], A.I.M. Jakaria Rahman [c], Raf Guns [c], and Tim C.E. Engels [c,d]

[a] KU Leuven, Dept. of Mathematics, Celestijnenlaan 200B, B-3001 Leuven, Belgium
[b] University of Antwerp (UA), IBW, B-2000 Antwerp, Belgium
[c] Centre for R&D Monitoring (ECOOM), Faculty of Social Sciences, University of Antwerp, Middelheimlaan 1, B-2020 Antwerp, Belgium
[d] Antwerp Maritime Academy, Noordkasteel Oost 6, B-2030 Antwerp, Belgium

**Abstract**

In a previous article (Rahman, Guns, Rousseau, and Engels, 2015) we described several approaches to determine the cognitive distance between two units. One of these approaches was based on what we called barycenters in N dimensions. The present note corrects this terminology and introduces the more adequate term 'similarity-adapted publication vectors'. Furthermore, we correct an error in normalization and explain the importance of scale invariance in determining cognitive distance. We also consider weighted cosine similarity as an alternative approach to determine cognitive (dis)similarity. Overall, we find that the three approaches (distance between barycenters, distance between similarity-adapted publication vectors, and weighted cosine similarity) yield very similar results.

**Keywords:** cognitive distances; barycenters; similarity matrices, similarity-adapted publication vectors; weighted cosine similarity.

**Introduction**

In Rahman, Guns, Rousseau, and Engels (2015) we investigated the following two questions:

i) How can we visualize the expertise of two entities (e.g., a research group and a panel) using publication data?

ii) How can we quantify the overlap of expertise between two entities (e.g., a research group and a panel) using publication data?

In this note we rephrase this second research question as: How can we obtain, using publication data, a meaningful distance or proximity measure which represents the cognitive distance or proximity between two units?

Entities or units were either experts, panels of experts, or research groups. In the case study these research groups were either research groups in physics or in chemistry working at the University of Antwerp, Belgium. For details we refer to Rahman et al. (2015).

Publications were assigned to Web of Science Subject Categories, in short WoS SCs. To answer the first research question we created overlay maps for a base map of science (Leydesdorff, & Rafols, 2009; Rafols, Porter, & Leydesdorff, 2010; Leydesdorff, Carley, & Rafols, 2013). In an overlay map, the original map – referred to as the base map – provides the location of each SC, whereas publication data is used to visualize the unit's number of publications for each SC. Numbers of publications were represented by the size of each node. These overlay maps provide an answer to the first research question.

In order to answer the second research question we consider three approaches: one based on using barycenters in two dimensions, a second one using similarity-adapted publication vectors and a third one using weighted cosine similarities. The last two are applied in N dimensions, where N denotes the total number of SCs. The first again uses overlay maps. We note that the base map can be considered to be universal and hence has nothing to do with the concrete data at hand. Each SC has a place on this map and hence can be characterized by the corresponding coordinates, denoted as $(L_{j,1}, L_{j,2})$, $j = 1, ..., N$. Now for each panel member and for each research group a barycenter is calculated and Euclidean distances between barycenters can be calculated. Coordinates of these barycenters (in 2 dimensions) are given as

$$C_1 = \frac{\sum_{j=1}^{N} m_j L_{j,1}}{T} \; ; \; C_2 = \frac{\sum_{j=1}^{N} m_j L_{j,2}}{T} \qquad (1)$$

where $m_j$ is the number of publications of the unit under investigation (panel member, research group) belonging to category j; this category j has coordinates $(L_{j,1}, L_{j,2})$ in the base map. T is the total number of publications of the unit under investigation. This is the barycenter approach as announced in the title of Rahman et al. (2015). In this way distances between entities, as represented by their barycenters, can be calculated leading to the quantitative results answering research question two. We want to point out that the term 'barycenter' as such, has no meaning. Any point can be the barycenter of infinitely many sets of points, possibly using sets of weights. We refer the reader to appendix A for a formal description of the notion of a barycenter.

We further note that in order to obtain meaningful distances these values must be scale-invariant. This means that the distance between points *P* and *Q* must be the same as the distance between the points *P* and *cQ*, where *c* is a strictly positive number. Indeed: the total output of a research group can be several orders of magnitude larger than that of one expert. This difference must not play a role in determining cognitive distances. The barycenter method explained above and in particular formulae (1) satisfy this requirement as multiplying all *m<sub>j</sub>s* with the same strictly positive factor leads to the same barycenter.

**How to calculate cognitive distance in N dimensions?**

*Similarity-adapted publication vectors (SAPV)*

In Rahman et al. (2015) we used a second quantitative approach, which was also referred to as a barycenter approach. In this approach, we used a matrix of similarity values between the WoS SCs as made available by Rafols, Porter, & Leydesdorff (2010) at http://www.leydesdorff.net/overlaytoolkit/map10.paj. These authors created a matrix of citing to cited SCs based on the Science Citation Index (SCI) and Social Sciences Citation Index (SSCI), which was cosine-normalized in the citing direction. The result is a symmetric N×N similarity matrix (here, N=224) which we denote by S = $(s_{ij})_{ij}$. Now each unit's publications are represented by an N-dimensional vector. Coordinates of these vectors are the number of publications in each WoS SC. Then we wrote in (Rahman et al., 2015):

> *A barycenter in N dimensions is determined as the point $C = (C_1, C_2, \ldots, C_N)$, where:*
>
> $$C_k = \frac{\sum_{j=1}^{N} m_j s_{jk}}{T} \qquad (2)$$
>
> *Here $s_{jk}$ denotes the k-th coordinate of WoS subject category j, $m_j$ is the number of publications in subject category j, and $T = \sum_{j=1}^{N} m_j$ is the total number of publications.*

Observe that we replaced (for clarity) *L = A* as used in Rahman et al. (2015) by S (the similarity matrix) and *M* (in the original publication) by *T* (the total number of publications of the unit under investigation). In Rahman et al. (2015) we provided concrete calculations of distances between these so-called barycenters of units. Although formula (1) and (2) look the same, their interpretation is different as will be explained.

The numerator of formula (2) is equal to the k-th coordinate of $S * M$, the multiplication of the similarity matrix S and the column matrix of publications $M = (m_j)_j$. We next include an example showing what is actually happening.

Let $N$ be 4. Assume that a unit has publication column $M = \begin{pmatrix} 4 \\ 1 \\ 0 \\ 0 \end{pmatrix}$.

Let $S = \begin{pmatrix} 1 & 0.1 & 0.3 & 0.8 \\ 0.1 & 1 & 0.2 & 0.1 \\ 0.3 & 0.2 & 1 & 0.6 \\ 0.8 & 0.1 & 0.6 & 1 \end{pmatrix}$ then $S * M = \begin{pmatrix} 4.1 \\ 1.4 \\ 1.4 \\ 3.3 \end{pmatrix}$.

Dividing by $T=5$ yields the vector $\frac{1}{5} \begin{pmatrix} 4.1 \\ 1.4 \\ 1.4 \\ 3.3 \end{pmatrix} = \begin{pmatrix} 0.82 \\ 0.28 \\ 0.28 \\ 0.66 \end{pmatrix}$.

Clearly, the resulting column vector is not a barycenter as it is not obtained as the result of a barycenter operation on a set of vectors.

The column vector $S * M/T$, resulting from the matrix product of matrix S and column vector M/T, can be interpreted as a pseudo-normalized (because normalization has been performed with respect to the sum of the coordinates of M and not with respect to $S * M$) publication vector that takes similarity into account. For this reason, we call $S * M$ a similarity-adapted publication vector, denoted as $M_{sa}$. In this example, this means that, for instance, the one publication in the second category also contributes (for 10%) to the publications in category 1. Although there is no original publication in category 4, we end up with a value 3.3 because category 1 and category 4 are very similar (80% similarity) and also the second category contributed. If we neglect similarity then S is the identity matrix and publication columns stay unchanged. We consider this method to be quite interesting as it provides a solution to the problem that WoS SCs have rather fuzzy borders.

The above example and discussion show that instead of normalizing before we applied the linear mapping, we should have done it afterward. Fig. 1 illustrates what we did in Rahman et al. (2015) and what we should have done.

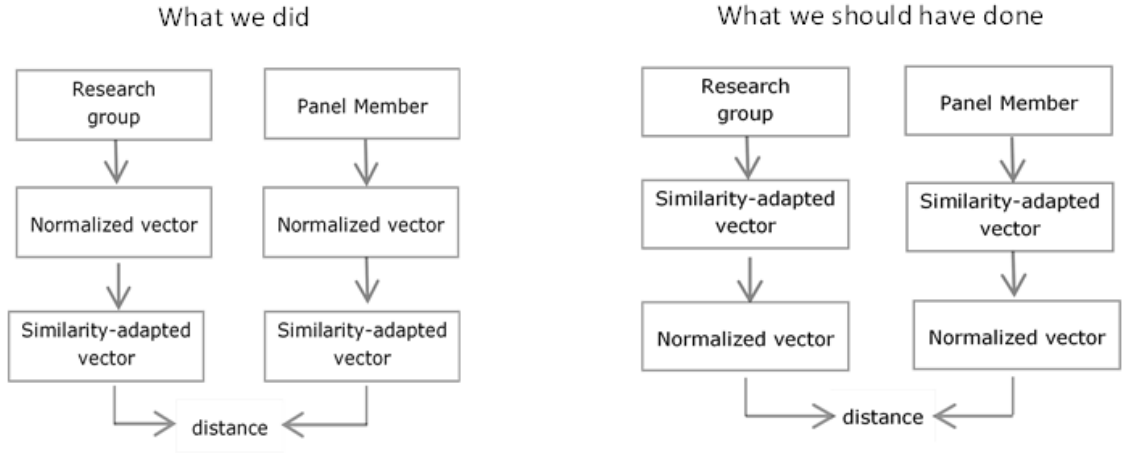

**Fig. 1.** Schematic comparison of the approach described by Rahman et al. (2015) and the improved approach presented in this note.

We note that instead of applying the approach shown in Fig. 1 to a research group and a panel member, we also apply it to panels and research groups as a whole.

The similarity-adapted publication vectors for which we calculated the distance in Rahman et al. (2015) are not normalized; hence the results obtained in our previous article are not scale-invariant. It suffices though, as shown on the right-hand side of Fig. 1 to reverse the order of the operations of normalization and matrix multiplication to end up with a scale-invariant distance. We will normalize by the $L_1$-norm of $M_{sa}$. Recall the four-dimensional example given above, where we found $S * M = \begin{pmatrix} 4.1 \\ 1.4 \\ 1.4 \\ 3.3 \end{pmatrix}$.

Rather than normalizing by 5, we need to divide by the sum of its coordinates. In this way, we obtain $\frac{1}{10.2} \begin{pmatrix} 4.1 \\ 1.4 \\ 1.4 \\ 3.3 \end{pmatrix} \approx \begin{pmatrix} 0.40 \\ 0.14 \\ 0.14 \\ 0.32 \end{pmatrix}$.

Hence, a similarity-adapted publication vector is determined as the vector $C = (C_1, C_2, \ldots, C_N)$, where:

$$C_k = \frac{\sum_{j=1}^{N} s_{kj} m_j}{\sum_{i=1}^{N} \sum_{j=1}^{N} s_{ij} m_j} \qquad (3)$$

where $s_{kj}$ denotes the similarity value between the $k$-th and the $j$-th WoS SC, and $m_j$ is the number of publications in WoS SC $j$. The numerator of Equation (3) is equal to the $k$-th element of $S * M$, the multiplication of the similarity matrix $S$ and the column matrix of publications $M = (m_j)_j$. The denominator is the $L_1$-norm of the unnormalized vector.

*Weighted cosine similarity*

Finally, we consider a weighted similarity method (generalized cosine similarity). The weighted similarity between panel member (PM) k and research group m, according to Zhou et al. (2012) is:

$$\frac{\sum_{i=1}^{N} M_i^k \left(\sum_{j=1}^{N} R_j^m s_{ji}\right)}{\sqrt{\left(\sum_{i=1}^{N} M_i^k \left(\sum_{j=1}^{N} M_j^k s_{ji}\right)\right) \cdot \left(\sum_{i=1}^{N} R_i^m \left(\sum_{j=1}^{N} R_j^m s_{ji}\right)\right)}} \qquad (4)$$

The numerator is nothing but the matrix multiplication: $(M^k)^t * S * R^m$, where $^t$ denotes matrix transposition, $S$ is the journal similarity matrix, $M^k$ denotes the column matrix of publications of panel member PMk and $R^m$ denotes the column matrix of publications of research group m. Similarly, the two products under the square root in the denominator are: $(M^k)^t * S * M^k$ and $(R^m)^t * S * R^m$. The result is the similarity between panel member PMk and research group m. This value is calculated for each panel member and each research group.

**Results**

By applying formula (3), we obtained a similarity-adapted publication vector for each entity in the similarity matrix. As in our previous paper, we can then calculate the Euclidean distance between different entities, e.g. individual research groups, individual panel members, all research groups together (Groups), and all panel members together (Panel).

Recall that the Euclidean distance between vectors a and b in $\mathbf{R}^N$ is given as:

$$d(a,b) = \sqrt{(a_1 - b_1)^2 + \cdots + (a_N - b_N)^2} \qquad (5)$$

The recalculation, in which normalization takes place after linear mapping, leads to the distances reported in Tables 1 and 2 for the cases of Chemistry and Physics. The group names have been standardized using the first four letters of the corresponding department, for example, CHEM-A for Chemistry research group A, PHYS-B for Physics research group B. The panel member names are standardized as PM1, PM2 etc.

**Table 1. Euclidean distances between similarity-adapted publication vectors of Chemistry individual research groups, panel members, groups and panel.**

| Groups | CHEM-A | CHEM-B | CHEM-C | CHEM-D | CHEM-E | CHEM-F | CHEM-G | CHEM-H | CHEM-I | CHEM-J | CHEM-K | CHEM-L |
|---|---|---|---|---|---|---|---|---|---|---|---|---|
| **Panel** | 0.047 | 0.054 | 0.047 | 0.073 | 0.033 | 0.080 | 0.078 | 0.067 | 0.056 | 0.046 | 0.089 | 0.101 | 0.049 |
| **PM 1** | 0.094 | 0.081 | 0.079 | 0.108 | 0.061 | 0.124 | 0.119 | 0.116 | 0.104 | 0.093 | 0.129 | 0.141 | 0.085 |
| **PM 2** | 0.043 | 0.082 | 0.074 | 0.079 | 0.054 | **0.036** | **0.032** | 0.055 | 0.046 | **0.036** | **0.075** | **0.071** | 0.070 |
| **PM 3** | 0.048 | 0.082 | 0.074 | 0.080 | 0.066 | 0.057 | 0.058 | 0.040 | **0.040** | 0.042 | 0.075 | 0.086 | 0.073 |
| **PM 4** | 0.069 | 0.106 | 0.099 | 0.104 | 0.085 | 0.064 | 0.070 | **0.027** | 0.063 | 0.071 | 0.085 | 0.094 | 0.091 |
| **PM 5** | 0.038 | **0.015** | **0.013** | **0.034** | 0.074 | 0.100 | 0.102 | 0.077 | 0.053 | 0.050 | 0.082 | 0.096 | **0.024** |
| **PM 6** | 0.082 | 0.093 | 0.087 | 0.111 | **0.025** | 0.085 | 0.080 | 0.096 | 0.090 | 0.080 | 0.113 | 0.116 | 0.088 |
| **PM 7** | 0.089 | 0.068 | 0.068 | 0.097 | 0.072 | 0.128 | 0.125 | 0.113 | 0.099 | 0.089 | 0.125 | 0.140 | 0.075 |

Shortest distances between a group and a panel member are underlined and printed in bold.

**Table 2. Euclidean distances between similarity-adapted publication vectors of Physics individual research groups, panel members, groups and panel.**

| Groups | PHYS-A | PHYS-B | PHYS-C | PHYS-D | PHYS-E | PHYS-F | PHYS-G | PHYS-H | PHYS-I |
|---|---|---|---|---|---|---|---|---|---|
| **Panel** | 0.021 | 0.154 | 0.018 | 0.030 | 0.255 | 0.028 | 0.109 | 0.094 | 0.021 | 0.112 |
| **PM 1** | 0.353 | 0.376 | 0.358 | 0.373 | **0.098** | 0.328 | 0.301 | 0.371 | 0.358 | 0.367 |
| **PM 2** | 0.044 | 0.172 | 0.019 | 0.038 | 0.272 | 0.054 | 0.127 | 0.115 | 0.019 | 0.133 |
| **PM 3** | 0.066 | 0.156 | 0.065 | 0.080 | 0.256 | 0.069 | **0.100** | 0.116 | 0.063 | 0.111 |
| **PM 4** | 0.042 | **0.144** | 0.060 | 0.039 | 0.271 | 0.051 | 0.129 | **0.066** | 0.063 | **0.103** |
| **PM 5** | 0.027 | 0.157 | 0.023 | **0.016** | 0.271 | 0.044 | 0.125 | 0.095 | 0.027 | 0.115 |
| **PM 6** | 0.032 | 0.165 | **0.012** | 0.035 | 0.258 | **0.037** | 0.111 | 0.106 | **0.015** | 0.125 |

Shortest distances between a group and a panel member are underlined and printed in bold.

Tables 1 and 2 replace, respectively, Table 1 and Table 3 of the supplementary online material, (part 2) of Rahman et al. (2015). In this connection, any reference to N dimensions and in particular to barycenters in N dimensions, and Figures 3.1.1 to 3.1.21 and 4.1.1 to 4.1.17 in the supplementary online material (part 3) should be ignored in Rahman et al. (2015).

By applying formula (4), we obtained the weighted cosine similarity (WCS) between panel members and the research groups in Chemistry (Table 3) and Physics (Table 4).

**Table 3. Weighted cosine similarity between Chemistry individual research groups, panel members, groups and panel.**

| Groups | CHEM-A | CHEM-B | CHEM-C | CHEM-D | CHEM-E | CHEM-F | CHEM-G | CHEM-H | CHEM-I | CHEM-J | CHEM-K | CHEM-L |
|---|---|---|---|---|---|---|---|---|---|---|---|---|
| **Panel** | 0.816 | 0.828 | 0.815 | 0.635 | 0.934 | 0.688 | 0.663 | 0.664 | 0.687 | 0.750 | 0.532 | 0.395 | 0.804 |
| **PM1** | 0.607 | 0.709 | 0.667 | 0.445 | 0.922 | 0.469 | 0.449 | 0.395 | 0.440 | 0.507 | 0.323 | 0.273 | 0.661 |
| **PM2** | 0.871 | 0.67 | 0.713 | 0.726 | 0.675 | **0.914** | **0.945** | 0.837 | 0.847 | **0.947** | 0.703 | **0.527** | 0.713 |
| **PM3** | 0.806 | 0.594 | 0.655 | 0.673 | 0.569 | 0.839 | 0.831 | 0.866 | **0.88** | 0.894 | **0.711** | 0.403 | 0.604 |
| **PM4** | 0.709 | 0.459 | 0.517 | 0.504 | 0.484 | 0.781 | 0.777 | **0.951** | 0.758 | 0.769 | 0.626 | 0.315 | 0.549 |
| **PM5** | 0.901 | **0.983** | **0.990** | **0.842** | 0.669 | 0.581 | 0.475 | 0.614 | 0.747 | 0.758 | 0.573 | 0.512 | **0.933** |
| **PM6** | 0.570 | 0.613 | 0.600 | 0.377 | **0.973** | 0.545 | 0.519 | 0.391 | 0.41 | 0.484 | 0.294 | 0.280 | 0.603 |
| **PM7** | 0.65 | 0.758 | 0.713 | 0.503 | 0.850 | 0.460 | 0.439 | 0.440 | 0.494 | 0.550 | 0.373 | 0.290 | 0.700 |

The highest similarity between a group and a panel member is underlined and printed in bold.

**Table 4. Weighted cosine similarity between Physics individual research groups, panel members, groups and panel.**

|      | Groups | PHYS-A | PHYS-B | PHYS-C | PHYS-D | PHYS-E | PHYS-F | PHYS-G | PHYS-H | PHYS-I |
|------|--------|--------|--------|--------|--------|--------|--------|--------|--------|--------|
| Panel | 0.988 | 0.196 | 0.97 | 0.939 | 0.324 | 0.933 | 0.618 | 0.673 | 0.976 | 0.57 |
| PM1 | 0.25 | 0.03 | 0.155 | 0.043 | **<u>0.996</u>** | 0.561 | 0.508 | 0.028 | 0.154 | 0.052 |
| PM2 | 0.939 | 0.151 | 0.982 | 0.92 | 0.127 | 0.806 | 0.513 | 0.543 | 0.977 | 0.497 |
| PM3 | 0.712 | **<u>0.22</u>** | 0.714 | 0.625 | 0.211 | 0.668 | 0.526 | 0.44 | 0.762 | 0.544 |
| PM4 | 0.804 | 0.182 | 0.729 | 0.829 | 0.129 | 0.757 | 0.436 | **<u>0.895</u>** | 0.741 | 0.479 |
| PM5 | 0.974 | 0.182 | 0.965 | **<u>0.986</u>** | 0.158 | 0.852 | 0.475 | 0.656 | 0.957 | **<u>0.567</u>** |
| PM6 | 0.979 | 0.164 | **<u>0.989</u>** | 0.93 | 0.272 | **<u>0.903</u>** | **<u>0.643</u>** | 0.631 | **<u>0.985</u>** | 0.516 |

The highest similarity between a group and a panel member is underlined and printed in bold.

We calculated the Pearson correlation coefficient (r) and the Spearman rank correlation coefficient (ρ) between distances based on the three approaches. These calculations are based on all distances between research groups and individual panel members. Results are shown in Tables 5 and 6. We recall that the 2-dimensional barycenter approach is based on a Kamada-Kawai map. Since the barycenter and SAPV approaches are distance-based rather than similarity-based, we use 1 – WCS as values to obtain dissimilarity values: weighted cosine dissimilarity, denoted as WCD, which can more easily be compared with the other two.

**Table 5. Pearson and Spearman correlation on all cognitive distances between Chemistry individual research groups and individual panel members.**

| Pearson / Spearman | Barycenter | SAPV | WCD |
|---|---|---|---|
| Barycenter | 1.00 | 0.75 | 0.64 |
| SAPV | 0.72 | 1.00 | 0.92 |
| WCD | 0.62 | 0.93 | 1.00 |

In the table 5, the upper triangle refers to Pearson correlations while the lower triangle refers to Spearman correlation values.

**Table 6. Pearson and Spearman correlation on all cognitive distances between Physics individual research groups and individual panel members.**

| Pearson / Spearman | Barycenter | SAPV | WCD |
|---|---|---|---|
| Barycenter | 1.00 | 0.29 (0.87) | 0.60 (0.89) |
| SAPV | 0.64 (0.94) | 1.00 | 0.86 (0.97) |
| WCD | 0.71 (0.91) | 0.94 (0.97) | 1.00 |

In the table 6, the upper triangle refers to Pearson correlations while the lower triangle refers to Spearman correlation values. Values between brackets are correlations calculated after removal of PHYS-D and PM1. Table 5 and 6 show that all correlations are high or moderately high except the Pearson correlation between the barycenter method and SAPV in the case of Physics. Fig. 2 illustrates what happened.

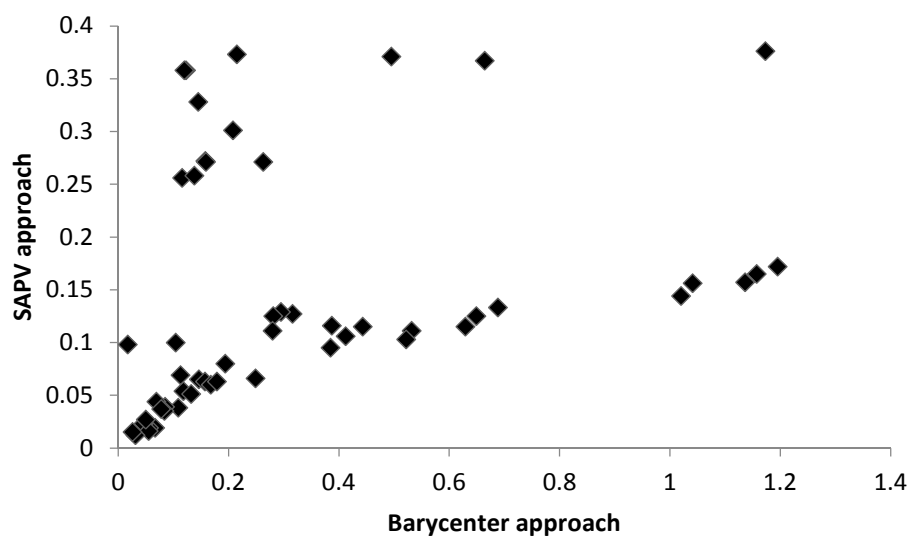

**Fig.2**. Scatter plot of the cognitive distances between groups and individual panel members for the barycenter and SAPV approaches in the Physics department.

This low Pearson correlation is due to the 13 points (including two times two points that overlap) in the upper half of Fig. 2. All these points correspond to distances involving research group PHYS-D and PM1 (but not both). This group and panel member are active in the same field (Physics, Particles & Fields) and have different scientific interests than the other groups or panel members: 99.1% of PM1's publications belong to the SC *Physics, Particles and Fields*, while for PHYS-D, this SC covers 83.6% of its publications. Moreover, their publications cover only four (117 publications) and seven (269 publications) WoS SCs respectively while other panel members cover 12 to 26 WoS SCs, and other research groups 26 to 50 SCs. Fig. 3 presents the same data as Fig. 2, but leaves out distances involving PHYS-D and PM1. In this case, r = 0.87 and ρ = 0.94. These values can also be seen in Table 6, where all values between brackets refer to correlations calculated without PHYS-D and PM1.

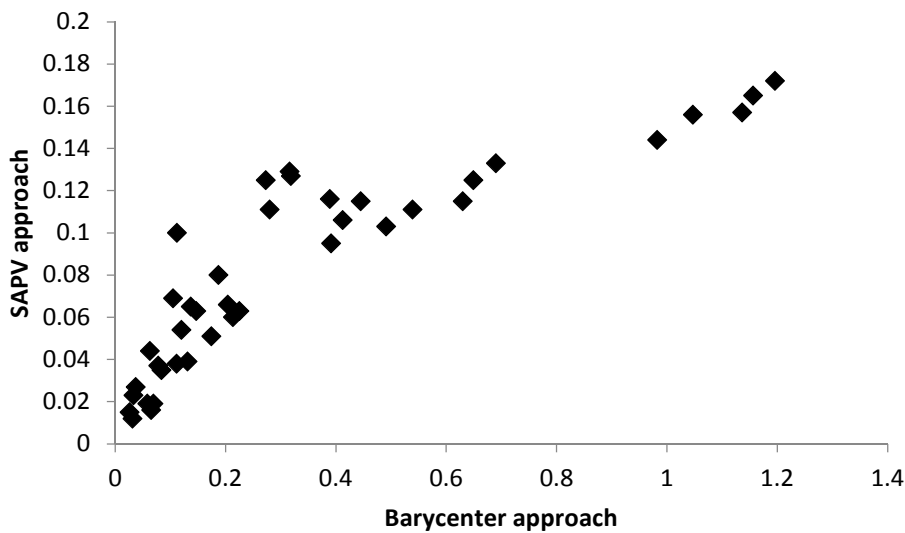

**Fig. 3**. Scatter plot of the cognitive distances between groups and individual panel members between barycenter and SAPV approaches in the Physics department excluding PHYS-D and PM1.

**Comparison between the three approaches**

In order to facilitate a comparison between the three approaches we included the results for the barycenter approach, as published in Rahman et al. (2015), in Appendix B. A comparison would be easy if a gold standard would exist. Clearly, it does not but we used the labor division decided upon by the panel chair as a proxy. Assuming that the panel chair has correct opinion on the expertise of the panel members and of the research group (remember that they never collaborated with them and neither do they belong to the same university, so this assumption does not necessarily always hold in practice), a perfect method would always rank this main assessor first.

Table 7 shows the research groups and the corresponding main assessor. For each of the three approaches, the three top-ranked panel members are shown. We have to point out two extra problems for chemistry. The first is that although PM7 was indicated as the main assessor for CHEM-C, PM3 thought himself closest to this research group. The second problem was that PM3 was indicated as main assessor of CHEM-F but he himself doubted if he could assess this group as an expert.

**Table 7. Top ranked panel members according to three approaches.**

| Research group | Main assessor | Barycenter | | | Similarity-adapted publication vector | | | Weighted cosine similarity | | |
|---|---|---|---|---|---|---|---|---|---|---|
| | | 3 | 2 | 1 | 3 | 2 | 1 | 3 | 2 | 1 |
| PHYS-A | PM3 | PM 4 | **PM 3** | PM 5 | PM 4 | **PM 3** | PM 5 | **PM 3** | PM 4 | PM 5 |
| PHYS-B | PM2 | PM 6 | PM 5 | **PM 2** | PM 6 | **PM 2** | PM 5 | PM 6 | **PM 2** | PM 5 |
| PHYS-C | PM5 | **PM 5** | PM 6 | PM 2 | **PM 5** | PM 6 | PM 2 | **PM 5** | PM 6 | PM 2 |
| PHYS-D | PM1 | **PM 1** | PM 3 | PM 6 | **PM 1** | PM 3 | PM 6 | **PM 1** | PM 6 | PM 3 |
| PHYS-E | PM4 | PM 5 | PM 6 | PM 3 | PM 6 | PM 5 | **PM 4** | PM 6 | PM 5 | PM 2 |
| PHYS-F | PM1 | PM 3 | **PM 1** | PM 5 | PM 3 | PM 6 | PM 5 | PM 6 | PM 3 | PM 2 |
| PHYS-G | PM4 | **PM 4** | PM 3 | PM 5 | **PM 4** | PM 5 | PM 6 | **PM 4** | PM 5 | PM 6 |
| PHYS-H | PM6 | **PM 6** | PM 5 | PM 2 | **PM 6** | PM 2 | PM 5 | **PM 6** | PM 2 | PM 5 |
| PHYS-I | PM3 | PM 4 | **PM 3** | PM 5 | PM 4 | **PM 3** | PM 5 | PM 5 | **PM 3** | PM 6 |
| **Score** | | | | **19** | | | **19** | | | **19** |
| | | | | | | | | | | |
| CHEM-A | PM6 | PM 5 | PM 3 | PM 7 | PM 5 | PM 7 | PM 1 | PM 5 | PM 7 | PM 1 |
| CHEM-B | PM5 | **PM 5** | PM 7 | PM 1 | **PM 5** | PM 7 | PM 2 | **PM 5** | PM 2 | PM 7 |
| CHEM-C | PM7 | PM 5 | PM 3 | **PM 7** | PM 5 | PM 2 | PM 3 | PM 5 | PM 2 | PM 3 |
| CHEM-C | PM3 | PM 5 | **PM 3** | PM 7 | PM 5 | PM 2 | **PM 3** | PM 5 | PM 2 | **PM 3** |
| CHEM-D | PM2 | PM 6 | PM 4 | PM 3 | PM 6 | **PM 2** | PM 1 | PM 6 | PM 1 | PM 7 |
| CHEM-E | PM2 | **PM 2** | PM 4 | PM 6 | **PM 2** | PM 3 | PM 4 | **PM 2** | PM 3 | PM 4 |
| CHEM-F | PM3 | PM 2 | PM 6 | PM 4 | PM 2 | **PM 3** | PM 4 | PM 2 | **PM 3** | PM 4 |
| CHEM-G | PM3 | **PM 3** | PM 4 | PM 2 | PM 4 | **PM 3** | PM 2 | PM 4 | **PM 3** | PM 2 |
| CHEM-H | PM5 | PM 4 | PM 2 | PM 3 | PM 3 | PM 2 | **PM 5** | PM 3 | PM 2 | PM 4 |
| CHEM-I | PM4 | PM 3 | **PM 4** | PM 5 | PM 2 | PM 3 | PM 5 | PM 2 | PM 3 | **PM 4** |
| CHEM-J | PM4 | **PM 4** | PM 2 | PM 3 | PM 2 | PM 3 | PM 5 | PM 3 | PM 2 | **PM 4** |
| CHEM-K | PM6 | PM 2 | PM 4 | PM 3 | PM 2 | PM 3 | PM 4 | PM 2 | PM 5 | PM 3 |
| CHEM-L | PM1 | PM 5 | PM 7 | **PM 1** | PM 5 | PM 2 | PM 3 | PM 5 | PM 2 | PM 7 |
| **Score with PM7 assigned to CHEM-C** | | | | **16** | | | **13** | | | **12** |
| **Score with PM3 assigned to CHEM-C** | | | | **17** | | | **14** | | | **13** |

We note that for some research groups the three approaches and the chosen assessor coincide. This perfect result was attained for PHYS-C, PHYS-D, PHYS-G, PHYS-H, CHEM-B and CHEM-E. For some other groups no approach lead to the chosen assessor, not even as second ranked one. This is the case for PHYS-E, CHEM-A, CHEM-H, CHEM-K and CHEM-L. In all these cases, the results for the three approaches agree to a very large extent. A possible explanation might simply be that the panel chair had no correct opinion on the expertise of the panel members and/or research groups. In the case of Chemistry where the suggested labor division was partly contested by PM3, that panel member seems indeed closer than PM7, raising the score in all three approaches. Yet another panel member, PM5, is identified in all three approaches as the closest to CHEM-C.

If the first ranked panel member was the main assessor, a score of 3 was assigned. The second ranked panel member corresponds to a score of 2 and the third ranked one to a score of 1. For lower ranked panel members, no score was assigned. The sum of all scores for a given approach in a given discipline yields the scores shown in Table 5. In the case of Physics the three approaches score equally. In the case of Chemistry, barycenters score best, followed by similarity-adapted publication vectors and, finally, weighted cosine similarity. Because the differences between the scores are so low, this ranking should not be generalized. In future research we intend to compare different methods using more data from different disciplines.

Before deciding on a choice between the methods presented in this contribution we need to point out that there is a technical problem with the generalized cosine similarity approach. Indeed, this method uses an inner product of the form $< x, Sy >$, see Zhou et al. (2012). Yet, this expression is only an inner product, i.e. satisfies all axioms for an inner product, if the similarity matrix S is positive definite. This is the case for the specific matrix S used in (Zhou et al., 2012) and also here, but this does not have to be the case in general (more details are given in (Zhou et al., 2012). For this reason, we think that the generalized cosine similarity method should not be used, as it does not always have good mathematical properties.

Finally, as the barycenter approach and the approach based on similarity-adapted publication vectors seem to lead to comparable results, we express a slight preference for the latter method. Indeed, this method uses distances in the original N-dimensional space, while the barycenter approach results from a projection into two dimensions. This projection can be performed in many different ways (we made the calculation in a Kamada-Kawai map) and certainly contains some distortions with respect to the N-dimensional one.

**Conclusion**

In a previous article, we included calculations derived from so-called barycenters of units in N dimensions. In retrospect, we admit that referring to these points as 'barycenters' was a misnomer. In this note we showed that, besides using barycenters in a two-dimensional base map it is possible to derive cognitive distances in N dimensions using similarity-adapted publication vectors or weighted cosine similarity. Of course, other approaches are also possible, such as the one proposed by Wang and Sandström (2015), which is based on bibliographic coupling and topic modelling.

As pointed out in this note, calculating cognitive distances between units should be scale invariant. Barycenters in a two-dimensional base map satisfy this requirement. We note though that distances in a 2D-map are artificial, hence only comparisons between distances and not their absolute values have meaning. Proper normalization in N dimensions also leads to scale invariant distances. Distances between normalized similarity-adapted publication vectors in N-dimensions are probably less distorted and hence more meaningful. Hence, our preliminary (as more tests should be done) choice goes to the approach based on similarity-adapted publication vectors (SAPV).

**Acknowledgments**

This investigation has been made possible by the financial support of the Flemish government to ECOOM. The opinions in the paper are the authors' and not necessarily those of the government. We thank the reviewers for constructive remarks.

**Appendix A. Barycenters**

A barycenter is the result of an operation performed on a set of vectors. Let $(X_n)_{n=1,\ldots,k}$ be a set of vectors in m-dimensional space, $\mathbf{R}^m$. Then its barycenter $B_X$ is the result of the following mapping:

$$B: (\mathbf{R}^m)^k \to R^m : (X_n)_{n=1,\ldots,k} \to B_X = \frac{1}{k}\sum_{n=1}^{k} X_n \qquad (6)$$

An example: let m = 2, k = 4 and $X_1 = \binom{0}{0}, X_2 = \binom{0}{1}, X_3 = \binom{2}{1}$ and $X_4 = \binom{2}{0}$. Then the barycenter of this set of four vectors is: $\frac{1}{4}\left(\binom{0}{0} + \binom{0}{1} + \binom{2}{1} + \binom{2}{0}\right) = \frac{1}{4}\binom{4}{2} = \binom{1}{0.5}$. This is the standard barycenter of the set of vertices $X_1$, $X_2$, $X_3$ and $X_4$ of a rectangle in the plane. More generally, one may assign a positive weight to each vector. If $m_n$ is the weight assigned to vector $X_n$ then the (generalized) barycenter (or center of gravity) is the result of the following mapping:

$$B: (\mathbf{R}^+, R^m)^k \to R^m : (m_n, X_n)_{n=1,\ldots,k} \to B_x = \frac{1}{T}\sum_{n=1}^{k} m_n X_n$$

where $T = \sum_{n=1}^{k} m_n$. If all weights are set equal to 1 then one recovers formula (6).

Clearly, any vector can be the barycenter of infinitely many sets of vectors and weights. This is the main reason why the term 'barycenter' has no meaning on its own. In an extremely formal way, one may even say that any vector X is the barycenter of its self, by taken the set of vectors equal to the singleton set {X} and weight equal to 1. Yet, this has no practical meaning whatsoever.

**Appendix B. Euclidean distances between barycenters**

**Table 8: Euclidean distances between barycenter of Chemistry individual research groups, panel members, groups and panel in the Kamada-Kawai map.**

|       | Groups | CHEM-A | CHEM-B | CHEM-C | CHEM-D | CHEM-E | CHEM-F | CHEM-G | CHEM-H | CHEM-I | CHEM-J | CHEM-K | CHEM-L |
|-------|--------|--------|--------|--------|--------|--------|--------|--------|--------|--------|--------|--------|--------|
| **Panel** | 0.113 | 0.151 | 0.118 | 0.186 | 0.106 | 0.274 | 0.347 | 0.114 | 0.173 | 0.061 | 0.265 | 0.357 | 0.141 |
| **PM 1** | 0.178 | 0.175 | 0.130 | 0.210 | 0.148 | 0.329 | 0.397 | 0.185 | 0.244 | 0.131 | 0.335 | 0.428 | 0.126 |
| **PM 2** | 0.189 | 0.305 | 0.299 | 0.330 | 0.136 | **_0.082_** | **_0.158_** | 0.137 | 0.111 | 0.172 | 0.194 | **_0.226_** | 0.337 |
| **PM 3** | 0.060 | 0.144 | 0.131 | 0.175 | 0.112 | 0.244 | 0.320 | **_0.052_** | 0.112 | **_0.003_** | 0.203 | 0.295 | 0.169 |
| **PM 4** | 0.115 | 0.229 | 0.225 | 0.255 | 0.110 | 0.156 | 0.232 | 0.063 | **_0.059_** | 0.099 | **_0.164_** | 0.231 | 0.265 |
| **PM 5** | 0.104 | **_0.047_** | **_0.022_** | **_0.083_** | 0.208 | 0.355 | 0.431 | 0.146 | 0.201 | 0.106 | 0.258 | 0.363 | **_0.069_** |
| **PM 6** | 0.208 | 0.289 | 0.261 | 0.323 | **_0.040_** | 0.171 | 0.229 | 0.170 | 0.196 | 0.156 | 0.302 | 0.361 | 0.282 |
| **PM 7** | 0.201 | 0.148 | 0.100 | 0.177 | 0.220 | 0.398 | 0.468 | 0.225 | 0.286 | 0.172 | 0.362 | 0.463 | 0.070 |

Shortest distances between a group and a panel member are underlined and printed in bold.

**Table 9. Euclidean distances between barycenter of Physics individual research groups, panel members, groups and panel in the Kamada-Kawai map.**

|       | Groups | PHYS-A | PHYS-B | PHYS-C | PHYS-D | PHYS-E | PHYS-F | PHYS-G | PHYS-H | PHYS-I |
|-------|--------|--------|--------|--------|--------|--------|--------|--------|--------|--------|
| Panel | 0.134 | 1.115 | 0.030 | 0.066 | 0.123 | 0.038 | 0.249 | 0.377 | 0.043 | 0.608 |
| PM 1 | 0.206 | 1.151 | 0.106 | 0.195 | **_0.013_** | 0.123 | 0.195 | 0.471 | 0.105 | 0.643 |
| PM 2 | 0.217 | 1.196 | 0.069 | 0.111 | 0.154 | 0.120 | 0.318 | 0.445 | 0.058 | 0.690 |
| PM 3 | 0.131 | 1.047 | 0.137 | 0.187 | 0.090 | 0.105 | **_0.112_** | 0.389 | 0.147 | 0.539 |
| PM 4 | 0.119 | **_0.982_** | 0.213 | 0.131 | 0.286 | 0.174 | 0.316 | **_0.204_** | 0.225 | **_0.491_** |
| PM 5 | 0.157 | 1.136 | 0.033 | **_0.065_** | 0.135 | **_0.063_** | 0.273 | 0.391 | 0.037 | 0.630 |
| PM 6 | 0.176 | 1.156 | **_0.031_** | 0.084 | 0.130 | 0.078 | 0.280 | 0.412 | **_0.026_** | 0.649 |

Shortest distances between a group and a panel member are underlined and printed in bold.